\documentclass[aps,pre]{revtex4}
\usepackage{latexsym}
\usepackage[percent]{overpic} 
\usepackage{float}
\usepackage[percent]{overpic} 
\DeclareGraphicsExtensions{.pdf,.png,.jpg,.eps}

\usepackage{color}
\usepackage{amssymb}
\usepackage{tikz}
\usepackage{amsmath}
\usetikzlibrary{shapes}

\begin{document}

\title{Cascading Failures in Interdependent Networks
  with Multiple Supply-Demand Links and Functionality Thresholds}

\author{M. A. Di Muro}
\affiliation{Instituto de Investigaciones F\'isicas de Mar del Plata
  (IFIMAR)-Departamento de F\'isica, Facultad de Ciencias Exactas y
  Naturales, Universidad Nacional de Mar del Plata-CONICET, Funes
  3350, (7600) Mar del Plata, Argentina.}
\email{mdimuro@mdp.edu.ar}

\author{L. D. Valdez}
\affiliation{Instituto de
  F\'isica Enrique Gaviola, CONICET, Ciudad Universitaria, 5000
  C\'ordoba, Argentina}
\affiliation{Facultad de Matem\'atica,
  Astronom\'ia, F\'isica y Computaci\'on, Universidad Nacional de
  C\'ordoba, 5000 C\'ordoba, Argentina}

\author{H. H. Arag\~{a}o R\^{e}go}
\affiliation{Departamento de F\'isica,
  Instituto Federal de Educa\c{c}\~{a}o, Ci\^{e}ncia e Tecnologia do
  Maranh\~{a}o, S\~{a}o Lu\'is, MA, 65030-005, Brazil}

\author{S. V. Buldyrev} \affiliation{Department of Physics, Yeshiva
  University, 500 West 185th Street, New York, New York 10033, USA}

\author{H. E. Stanley} \affiliation{Center for Polymer Studies, Boston
  University, Boston, Massachusetts 02215, USA}

\author{L. A. Braunstein} \affiliation{Instituto de Investigaciones
  F\'isicas de Mar del Plata (IFIMAR)-Departamento de F\'isica,
  Facultad de Ciencias Exactas y Naturales, Universidad Nacional de
  Mar del Plata-CONICET, Funes 3350, (7600) Mar del Plata, Argentina.}
\affiliation{Center for Polymer Studies, Boston University, Boston,
  Massachusetts 02215, USA}

\begin{abstract}
Various social, financial, biological and technological systems can be
modeled by interdependent networks. It has been assumed that in order
to remain functional, nodes in one network must receive the support
from nodes belonging to different networks. So far these models have
been limited to the case in which the failure propagates across
networks only if the nodes lose all their supply nodes. In this paper
we develop a more realistic model for two interdependent networks in
which each node has its own supply threshold, i.e., they need the
support of a minimum number of supply nodes to remain functional. In
addition, we analyze different conditions of internal node failure due
to disconnection from nodes within its own network. We show that
several local internal failure conditions lead to similar nontrivial
results. When there are no internal failures the model is equivalent
to a bipartite system, which can be useful to model a financial
market. We explore the rich behaviors of these models that include
discontinuous and continuous phase transitions. Using the generating
functions formalism, we analytically solve all the models in the limit
of infinitely large networks and find an excellent agreement with the
stochastic simulations.
\end{abstract}

\newcommand{\bcktrgle}{\raisebox{0.5pt}{\tikz{\node[draw,scale=0.35,regular polygon, regular polygon sides=3,fill=black,rotate=0](){};}}}
\newcommand{\rdbox}{\raisebox{0.5pt}{\tikz{\node[draw,scale=0.5,regular polygon, regular polygon sides=4,color=red,fill=red](){};}}}
\newcommand{\grdiam}{\raisebox{0pt}{\tikz{\node[draw,scale=0.4,diamond,color=green,fill=green](){};}}}
\newcommand{\invblutrgle}{\raisebox{0pt}{\tikz{\node[draw,scale=0.35,regular polygon, regular polygon sides=3,color=blue,fill=blue,rotate=180](){};}}}
\newcommand{\mgcirc}{\raisebox{0.5pt}{\tikz{\node[draw,scale=0.5,circle,color=magenta,fill=magenta](){};}}}
\newcommand{\bckcirc}{\raisebox{0.5pt}{\tikz{\node[draw,scale=0.5,circle,fill=black](){};}}}
\newcommand{\rddiam}{\raisebox{0pt}{\tikz{\node[draw,scale=0.4,diamond,color=red,fill=red](){};}}}
\newcommand{\blubox}{\raisebox{0.5pt}{\tikz{\node[draw,scale=0.5,regular polygon, regular polygon sides=4,color=blue,fill=blue](){};}}}
\newcommand{\grtrgle}{\raisebox{0.5pt}{\tikz{\node[draw,scale=0.35,regular polygon, regular polygon sides=3,color=green,fill=green,rotate=0](){};}}}
\newcommand{\unfbckcirc}{\raisebox{0.5pt}{\tikz{\node[draw,scale=0.5,circle,color=black,line width=0.7pt,fill=none](){};}}}
\newcommand{\unfrddiam}{\raisebox{0pt}{\tikz{\node[draw,scale=0.4,diamond,line width=0.7pt,color=red,fill=none](){};}}}
\newcommand{\unfblutrgle}{\raisebox{0.5pt}{\tikz{\node[draw,scale=0.35,regular polygon, regular polygon sides=3,line width=0.7pt,color=blue,fill=none,rotate=0](){};}}}
\newcommand{\unfmgbox}{\raisebox{0.5pt}{\tikz{\node[draw,scale=0.5,regular polygon, regular polygon sides=4,line width=0.7pt,color=magenta,fill=none](){};}}}
\newcommand{\unfrdtrgle}{\raisebox{0.5pt}{\tikz{\node[draw,scale=0.35,regular polygon, regular polygon sides=3,line width=0.7pt,color=red,fill=none,rotate=0](){};}}}
\newcommand{\ungrdiam}{\raisebox{0pt}{\tikz{\node[draw,scale=0.4,diamond,line width=0.7pt,color=green,fill=none](){};}}}
\newcommand{\unfblubox}{\raisebox{0.5pt}{\tikz{\node[draw,scale=0.5,regular polygon, regular polygon sides=4,line width=0.7pt,color=blue,fill=none](){};}}}
\definecolor{verdecito}{RGB}{135,190,69}
\definecolor{azu}{RGB}{0,160,227}
\definecolor{roj}{RGB}{229,50,36}
\definecolor{pinku}{RGB}{240,134,131}
\definecolor{sky}{RGB}{0,191,255}
\newcommand{\verdcirc}{\raisebox{0.5pt}{\tikz{\node[draw,scale=0.5,circle,color=verdecito,fill=verdecito](){};}}}
\newcommand{\azcirc}{\raisebox{0.5pt}{\tikz{\node[draw,scale=0.5,circle,color=azu,fill=azu](){};}}}
\newcommand{\rojcirc}{\raisebox{0.5pt}{\tikz{\node[draw,scale=0.5,circle,color=roj,fill=roj](){};}}}
\newcommand{\pkcirc}{\raisebox{0.5pt}{\tikz{\node[draw,scale=0.5,circle,color=pinku,fill=pinku](){};}}}
\newcommand{\bckline}{\raisebox{2pt}{\tikz{\draw[black,solid,line width=1pt](0,0) -- (5mm,0);}}}
\newcommand{\bludash}{\raisebox{2pt}{\tikz{\draw[blue,dashed,line width=1pt](0,0) -- (5mm,0);}}}
\newcommand{\rddotted}{\raisebox{2pt}{\tikz{\draw[red,dotted,line width=1.8pt](0,0) -- (5mm,0);}}}
\newcommand{\skylinex}{\raisebox{2pt}{\tikz{\draw[sky,dashdotted,line width=1pt](0,0) -- (5mm,0);}}}

\definecolor{indigo}{RGB}{75,0,130}
\definecolor{verd}{RGB}{0,139,0}
\definecolor{orange4}{RGB}{247,114,81}
\newcommand{\unfmgtrgle}{\raisebox{0.5pt}{\tikz{\node[draw,scale=0.35,regular polygon, regular polygon sides=3,line width=0.7pt,color=magenta,fill=none,rotate=0](){};}}}
\newcommand{\unfinvogtrgle}{\raisebox{0.5pt}{\tikz{\node[draw,scale=0.35,regular polygon, regular polygon sides=3,line width=0.7pt,color=orange,fill=none,rotate=180](){};}}}
\newcommand{\unfcybox}{\raisebox{0.5pt}{\tikz{\node[draw,scale=0.5,regular polygon, regular polygon sides=4,line width=0.7pt,color=cyan,fill=none](){};}}}
\newcommand{\magline}{\raisebox{2pt}{\tikz{\draw[magenta,dashed,line width=1pt](0,0) -- (5mm,0);}}}
\newcommand{\bluline}{\raisebox{2pt}{\tikz{\draw[blue,dashdotted,line width=1pt](0,0) -- (5mm,0);}}}
\newcommand{\orgline}{\raisebox{2pt}{\tikz{\draw[orange4,dotted,line width=1pt](0,0) -- (5mm,0);}}}
\newcommand{\indigline}{\raisebox{2pt}{\tikz{\draw[indigo,dashed,line width=1pt](0,0) -- (5mm,0);}}}
\newcommand{\skyline}{\raisebox{2pt}{\tikz{\draw[sky,dotted,line width=1pt](0,0) -- (5mm,0);}}}
\newcommand{\verdline}{\raisebox{2pt}{\tikz{\draw[verd,dashdotted,line width=1pt](0,0) -- (5mm,0);}}}
\newcommand{\reddotline}{\raisebox{2pt}{\tikz{\draw[red,dotted,line width=1pt](0,0) -- (5mm,0);}}}
\newcommand{\magdashdotline}{\raisebox{2pt}{\tikz{\draw[magenta,dashdotted,line width=1pt](0,0) -- (5mm,0);}}}
\newcommand{\bludotline}{\raisebox{2pt}{\tikz{\draw[blue,dotted,line width=1pt](0,0) -- (5mm,0);}}}
\newcommand{\orgdashline}{\raisebox{2pt}{\tikz{\draw[orange4,dashed,line width=1pt](0,0) -- (5mm,0);}}}

\flushbottom
\maketitle

\thispagestyle{empty}

\section*{Introduction}
Studying complex systems includes analyzing how the different components
of a given system interact with each other and how this interaction
affects the system's global colletive behavior. In recent years complex
network research has been a powerful tool for examining these systems,
and the initial research on isolated networks has yielded interesting
results \cite{Newman,Estrada_book_2011,Barrat_book_2013}.

A network is a graph composed of nodes that represent interacting
individuals, companies, or elements of an infrastructure. Node
interactions are represented by links or edges. Real-world systems
rarely work in isolation and often crucially depend on one another
\cite{Ros_01,report,Peerenboom,Rinaldi,Yagan,Kenett,Wang}.  Thus single-network
models have been extended to more general models of interacting
coupled networks, the study of which has greatly expanded our
understanding of real-world complex systems.  One intensive study of
these ``networks of networks'' has focused on the propagation of
failure among closely-related systems
\cite{Buldyrev2010,Buldyrev2011,Parshani,Shao,Huang,Gao2011,Gao2012,Li,Gao2012a,Bashan,Gao2013,Son_01,Bianconi2014,Bax_01,Valdez2013,Valdez2014}.
The great blackout of Italy in 2003 and the earthquake of Japan in
2011 were catastrophic events that demonstrated that breakdowns in
power grids strongly impact other systems such as communication and
transport networks, and that the failure of these networks in turn
accelerates the failure of the power grid. The propagation of these
``failure cascades'' has received wide study in recent years
\cite{Buldyrev2010,Buldyrev2011,Parshani,Shao,Huang,Gao2011,Gao2012,
  Li,Gao2012a,Bashan,Gao2013,DiMuro2016,DiMuroPRE}.

The simplest model of these systems consists of two interdependent
networks in which nodes in one network are connected by a single
bidirectional edge to nodes in a second network
\cite{Buldyrev2010}. In this model a node is functional (i) if it
belongs to the largest connected component (the ``giant component'')
in its own network (the internal rule of functionality) and (ii) if
its counterpart in the other network is also functional (the external
rule of functionality).  This original model has been extended to
include localized and targeted attacks
\cite{Yuan_2016,Dong_2013,Tani_2012,Huang,Zhao} and mitigation
\cite{Sch_01,Parshani,Ron_02,Valdez2014,Valdez2013} and recovery
strategies \cite{DiMuro2016,Hu_2016}. Recently it was found that the
giant component membership requirement can be replaced by a weaker
requirement of belonging to a cluster of a size larger than or equal
to a threshold $h^*$ \cite{DiMuroPRE}. Alternatively, a heterogeneous
k-core condition can be applied as an internal functionality condition
in which node $i$ is functional when at least $k^*_i$ nodes among its
$k_i$ immediate neighbors remain functional
\cite{Bax_2011,Cellai_13,Doro_2002,Doro_2008,Bax_2010}. In this model
the random failure of a critical fraction of nodes in an isolated
network leads to an abrupt collapse of this network.

Although the original interdependent network model expanded our
understanding of different coupled systems, the single-dependency
relationship between nodes in different networks does not accurately
represent what happens in real-world structures. A cascading failure
model of a network of networks with multiple dependency edges has been
applied to a scenario in which nodes fail only when they lose all their
support nodes in the other network \cite{Shao,Gao2012}, but nodes in complex
real-world systems can be so fragile that the loss of a single support
link can cause them to shut down. More generally, each node may require
a certain minimal number of supply links connected to the nodes in the
other network to remain functional.  In the world-wide economic system,
for example, banks and financial firms lend money to non-financial
companies who must pay the amount back with interest after a stated
period of time. If a single non-financial company becomes insolvent, the
bank that lent money to this company will likely not fail, but if the
number of companies that cannot pay back their loans is sufficiently
large, the possibility of bank failure becomes real. This resembles the
k-core process in a single network described above.

Here we model the process of cascading failure in a system of two
interdependent networks $A$ and $B$ in which nodes have multiple
connections or supply-demand links between networks. In the following,
network $X$ means either network $A$ or $B$. Each node $i$ in network
$X$ has $k_{sX,i}$ supply nodes in the other network that are connected
to node $i$ by supply links. This node remains functional at a
certain stage of the cascade of failures if the number of its functional
supply nodes in the other network remains greater or equal to its supply
threshold $k_{sX,i}^\ast \leq k_{sX,i}$. We call this the external
functionality condition. We assume that a supply threshold is predefined
for each node.

In principle, this model is non-trivial even if the survival of a node
in network $X$ does not directly depend on the internal connectivity of
network $X$.  In this case our model is equivalent to cascading failures
in a bipartite network composed of two sets of nodes $A$ and $B$
connected only by supply-demand links, i.e., these networks only have
external functionality. For generality, we add to the external
functionality condition an internal functionality condition that can be
one of the following: a node is functional (i) when it belongs to the
giant component of its network (``giant component rule''), (ii) when it belongs to a finite
component of size $h$ that survives with probability $1-q(h)$ (``mass rule''), and (iii)
when a node $i$ with internal connectivity $k_i$ has a number of
functional neighbors greater than or equal to $k_i^*$ (``$k$-core rule'').

We develop a theoretical model that is solved using the formalism of
generating functions. We present numerical solutions and compare them
with stochastic simulations. We find that for all internal rules of
functionality, increasing the $k^*_{sX}$ value increases system
vulnerability and often causes a discontinuous transition. For the
mass rule of internal functionality we find a continuous transition
for some parameter values. We also study the asymptotic limit of a large
number of supply links, and we find a relation between the critical
threshold of initial failure and the ratio $k^*_{sX}/k_{sX}$.

\section*{Model}\label{S.M}

\noindent
We assume that the system consists of two networks $A$ and $B$ with
internal degree distributions $P_A(k)$ and $P_B(k)$, respectively, where
$k$ is the degree of a node within its own network. Each node $i$ in
network $A$ is supplied by $k_{sA,i}$ supply links from nodes in network
$B$, and each node $j$ in network $B$ has $k_{sB,j}$ demand links that
act as supply links for nodes in network $A$. For simplicity we assume
that the demand links in network $A$ serve as supply links for nodes in
network $B$, and that supply links in network $A$ serve as demand links
for nodes in network $B$. Thus each supply-demand link is a
bidirectional link that connects a node in network $A$ with a node in
network $B$. If the internal degree of all nodes in networks $A$ and $B$
is zero, our model is equivalent to a bipartite network.  We assume that
the degree distribution of supply-demand links in network $A$ is
$P_{sA}(k)$ and the degree distribution of supply-demand links in
network $B$ is $P_{sB}(k)$.  In principle, some nodes may not have
supply links and still remain functional\cite{Parshani}. If this is the case,
$P_{sA}(0)>0$.

\begin{figure}[ht]
  \centering
  \includegraphics[width=300pt]{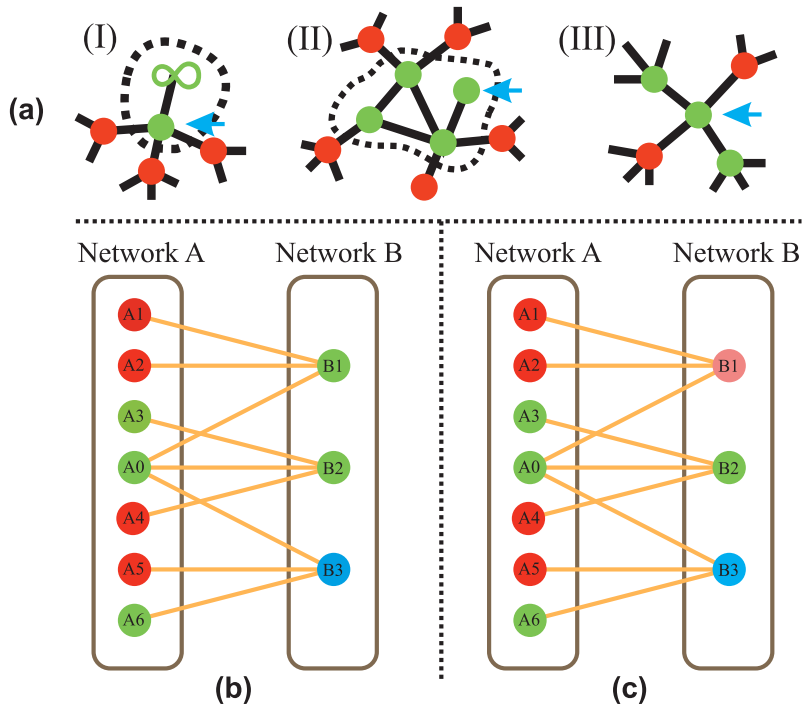}
\caption{Schematic of the rules of functionality of the model.  Black
  links represent internal connections and orange links the supplies
  between networks. The state of the nodes varies according to their
  color: functional nodes (\protect\verdcirc), nodes that do not fulfill
  the internal rule of functionality (\protect\azcirc) and nodes that
  fail due to the initial damage (\protect\rojcirc). In addition, we
  have nodes that externally fail because they do not get enough supply
  from the other network (\protect\pkcirc). In panel (a) we show the
  three internal rules of functionality for a node $i$ (marked with the
  blue arrow) to be functional: (I) it must be connected to the GC
  (represented by the $\infty$ symbol), (II) it must belong to a
  component of size $h$ which survives with probability $1-q(h)$ (in
  this case $q(4)<1$), or (III) it must have a number of neighbors equal
  to or greater than $k^* \equiv k^*_i$ (we show $k^*=2$ ). In panel (b)
  we show the external rule of functionality for $k_{s}^*=1$, and
  $k_{s}^*=2$ in panel (c). In these cases
  $P_{sA}(k)=P_{sB}(k)=\delta_{k,3}$, however, not all supplies are
  shown, nor are the internal connectivity links.}\label{scheme}
\end{figure}

The functionality of the nodes in both networks is related to their
connections within their own network, which we call the internal rule of
functionality. In addition, the state of the nodes also depends on the
supply demand links that connect both networks, which we call the
external rule of functionality.

We study three different internal rules of functionality:

\begin{itemize}\label{models}

\item[{(I)}] Model I (The ``giant component'' rule): nodes that belong
  to the giant component in their own network are functional.

\item[{(II)}] Model II (The ``finite component'' or ``mass'' rule): a
  finite component of size $h$ remains functional with a probability
  $1-q(h)$. If it fails, all of its nodes fail.  If it survives, all of
  its nodes remain functional.

\item[{(III)}] Model III (The ``k-core'' rule): a node $i$ with internal
  connectivity $k_i$ remains active if the number of its functional
  neighbors is greater than or equal to $k^*_i$.

\end{itemize}

\noindent
The external rule of functionality states that nodes in network $X$
must be connected with the other network through a number of
functional supply links greater than or equal to $k_{sX}^\ast$.

We call $k_{sX,i}^\ast$ the supply-demand functionality threshold of
node $i$, since in principle the threshold may be different for
different nodes. For conceptual simplicity, we assume that the supply
thresholds are predefined for each node by random selection from a
cumulative probability distribution $r_{sX}(j,k)=P(k_{sX}^\ast \leq j
\mid k_{sX}=k)$, where $P(\mid)$ is the conditional
probability. Alternatively, function $r_{sX}(j,k)$ can be understood as a probability
that a node with $k$ supply links remains functional if $j$ of its $k$
supply nodes in the other network remains functional.

For example, in the case of a uniform supply threshold $k_{sX}^\ast=m$
where $m$ is a constant, the distribution $r_{sX}$ is a step function,
i.e., $r_{sX}(j,k)=0$ for $j<m$ and $r_{sX}(j,k)=1$ for $j \geq
m$. Another option is linear: $r_{sX}(j,k)=j/k$. For autonomous nodes
that can survive without any functional supply nodes in the other
network, $k_{sX}^\ast=0$. This case is included in the general scheme if
we assume that $r_{sX}(0,k_{sX})>0$.

Figure~\ref{scheme}(a) shows a schematic of the internal rules of
functionality, and Figs.~\ref{scheme}(b) and \ref{scheme}(c) show a
schematic of the external rules of functionality.  In each network
green nodes are functional, i.e., they satisfy both internal and
external conditions of functionality. Red nodes are affected by the
initial failure, blue nodes do not satisfy internal conditions of
functionality and pink nodes do not satisfy external conditions of
functionality.  Internal links are black, and supply links are
orange. Here we use $P_{sA}(k)=P_{sB}(k)=\delta_{k,3}$, but for
simplicity in Figs.~\ref{scheme}(b) and \ref{scheme}(c) we omit the
internal links and some of the supply-demand links in network $A$. For
example, in Fig.~\ref{scheme}(b) node $A3$ has two additional supply
nodes from network $B$ that are not shown. Figure~\ref{scheme}(b)
shows the case $k_{s,i}^*=1$ for all $i$. Note that since all nodes in
network $B$ receive supplies from functional node $A0$ they are
unaffected when other nodes in network $A$ fail. On the other hand,
Fig.~\ref{scheme}(c) shows that when $k_{s,i}^*=2$ all nodes must have
two functional supply nodes from the other network to remain
functional. Nodes $B2$ and $B3$ are connected to $A0$, receive
supplies from functioning nodes $A3$ and $A6$, respectively, and
remain active. On the other hand, because node $B1$ is only supported
by node $A0$, it fails, as indicated by the pink color.

\subsection*{Theoretical approach}\label{T.a}

\noindent
We construct a system of two randomly connected networks in which
connectivity links within each network follow degree distributions
$P_A(k)$ and $P_B(k)$ and supply-demand links between the networks
follow distributions $P_{sA}(k)$ and $P_{sB}(k)$. For this system we
achieve a theoretical solution within the limit of a large number of
nodes, $N_A$ and $N_B$, where $N_A$ and $N_B$ are the number of nodes
in networks $A$ and $B$, respectively.  The bidirectionality of the
supply-demand links requires that relation $N_A\langle
k\rangle_{sA}=N_B\langle k\rangle_{sB}$ is satisfied, where $\langle
k\rangle_{sA}$ and $\langle k\rangle_{sB}$ are the average degrees of
the supply links in networks A and B respectively.

When we randomly remove a fraction $1-y_X$ of nodes from network $X$,
the remaining fraction of active nodes $\mu_X$ for an isolated network
$X$ is determined by which internal functionality rule is followed. It can be
expressed in the closed-form expression $\mu_X=y_X g_X(y_X)$, where
$g_X(y_X)\leq 1$ is an exacerbation factor that takes into account
additional node failures triggered by the random removal of a fraction
of $1-y_X$ nodes.  The explicit form of this factor is determined by the
internal functionality rules of the model.  The Supplementary
Information presents equations for $g_X$ for Rules I, II, and III (see
Supplementary Information: section {\it Explicit form of the
  functionality rules}). For example, for a bipartite network
$g_X(y_X)=1$.

The cascading process begins with a random failure in network $A$. This
failure causes an additional loss of nodes determined by the
exacerbation factor. This event triggers a cascade in which failure is
transmitted back and forth between networks $A$ and $B$ through the
supply-demand links, and this further decreases the fraction of
functional nodes. The external functionality rule states that node $i$
with $k_{s,i}$ supply-demand links must have $k_{s,i}^\ast$ or more
nodes to remain functional, similar to k-core percolation.

External functionality failure is similar to heterogeneous k-core
percolation \cite{Cellai_13}. To describe this failure due to a lack of
supply between networks $A$ and $B$, we introduce the functions
$W_{sA}(x),W_{sB}(x)$ and $Z_{sA}(x),Z_{sB}(x)$, which are the k-core
generating functions of the degree distribution and the excess degree
distribution of the supply-demand links in networks $A$ and $B$,
respectively.  These functions depend on the degree distributions
$P_{sA}$ and $P_{sB}$ of supply-demand links and the distribution of the
thresholds $r_{sA}(j,k)$ and $r_{sB}(j,k)$ of the supply-demand links in
networks $A$ and $B$,

\begin{equation}\label{WWs}
W_{sX}(\beta)=\sum_{k=0}^\infty P_{sX}(k)\sum_{j=0}^k\binom{k}{j}r_{sX}(j,k)\beta^j(1-\beta)^{k-j}
\end{equation}
and
\begin{equation}\label{ZZs}
Z_{sX}(\beta)=\sum_{k=0}^\infty \frac{k P_{sX}(k)}{\langle k_s\rangle_{X}}\sum_{j=0}^{k-1}\binom{k-1}{j}r_{sX}(j+1,k)\beta^{j}(1-\beta)^{k-j-1}, 
\end{equation}
where $\langle k_s\rangle_{X}$ is the average number of supply links
per node in network $X$. In this context $\beta$ is the probability that
a functional node will be selected. Similar formulas were derived in
Ref.~\cite{Gleeson} for a variant of the Watts opinion model
\cite{Watts}.

We next examine a theoretical approach to the temporal evolution of the
cascading process. As explained above, initially a randomly selected
fraction $1-p$ of nodes fails in network $A$. Then the surviving
fraction of nodes in network $A$ in this first stage of the cascade is
$\mu_{A,1}=pg_A(p)$. We introduce a new parameter $f_B$, which is the
probability of randomly choosing a supply link that is connected to a
functional node in the other network. When a node fails, all its demand
links also fail. Thus $f_{B,1}=\mu_{A,1}$

After applying the external functionality rule to network $B$, the
fraction of nodes that fulfill the conditions is given by
$y_{B,1}=W_{sB}(f_{B,1})$. Because there are additional disconnected
nodes in network $B$ given by the exacerbation factor $g_B$, the number
of functional nodes in network $B$ at the first stage of the cascade is
$\mu_{B,1}=y_{B,1}g_B(y_{B,1})$. In the second stage of the cascade we
cannot apply the same rules to obtain $\mu_{A,2}$, because $f_{A,2}\neq
\mu_{B,1}$. If, for example, a supply-demand link connects nodes $i$ in
$A$ and $j$ in $B$, then the probability that this link is active
depends on how many other links belonging to nodes $i$ or $j$ are
active. Thus the fraction of surviving links at this step is
$f_{A,2}=Z_{sB}(f_{B,1})g_B(y_{B,1})$.

Thus the recursion relations for the stages $n>1$ are
\begin{eqnarray}\label{Eq.f}
  f_{A,n} &=& Z_{sB}(f_{B,n-1})\;g_B(y_{B,n-1}) \nonumber;\\
  f_{B,n} &=& p\;Z_{sA}(f_{A,n})\;g_A(y_{A,n}),
\end{eqnarray}
where 
\begin{eqnarray}\label{Eq.y}
  y_{A,n} &=& p\;W_{sA}(f_{A,n}) \nonumber;\\
  y_{B,n} &=& W_{sB}(f_{B,n})
\end{eqnarray} 
are the fractions of nodes that satisfy the external rule of
functionality, i.e., {\it randomly\/} removing a fraction of
$1-y_{X,n}$ nodes leaves the same number of functional nodes as in
stage $n$ of the cascade. The fractions of functional nodes at stage
$n$ of the cascade are
\begin{eqnarray}\label{Eq.mu}
\mu_{A,n} &=& y_{A,n}\;g_A(y_{A,n}) \nonumber;\\
\mu_{B,n} &=& y_{B,n}\;g_B(y_{B,n}).
\end{eqnarray} 
The process begins with $f_{A,1}=1$ and $y_{A,1}=p$, which is equivalent
to an initial random failure on network $A$.
\section{Results} 

\noindent
We next present these theoretical results using several simple examples
and verifying them with stochastic simulations.

\begin{figure}[ht] 
\centering
\includegraphics[width=\linewidth]{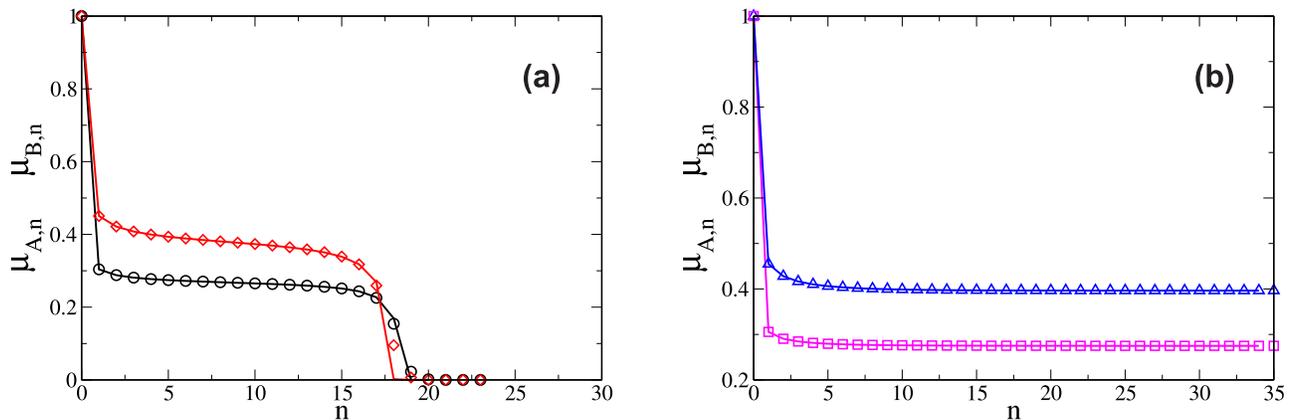}
\caption{Temporal evolution, close to the critical threshold, of the
  giant component $\mu_A(n)$ and $\mu_B(n)$ of networks $A$ and $B$,
  when both are random regular (RR) networks with delta degree
  distribution $P_X(k)=\delta_{k,5}$, with $X=A,B$.  The degree
  distributions of supply links are also delta-distributions with
  $P_{sA}(k)=P_{sB}(k)=\delta_{k,5}$ and $k_{s}^*=2$. The critical
  threshold for this system is $p_c=0.381$. (a) $p=0.38$, (b)
  $p=0.381$. Network A (\protect\unfbckcirc,\;\protect\unfmgbox),
  Network B (\protect\unfrddiam,\;\protect\unfblutrgle). The dashed
  lines are the results from the equations and the symbols are the
  results from the stochastic simulations.}\label{Pinf_n}
\end{figure}

To test the validity of the equations, Fig.~\ref{Pinf_n} shows the
temporal evolution of the order parameter of networks $A$ and $B$ close
to the critical threshold $p_c$, computed using the equations and
stochastic simulations when the giant component functionality rule is
applied (see Supplementary Information: subsections {\it Giant
  Component} and {\it Numerical Solution for the threshold $p_c$}). Note
that the plots show the simulation results are in total agreement with
the theoretical results.

\begin{figure}[H] 
  \centering
    \includegraphics[width=\linewidth]{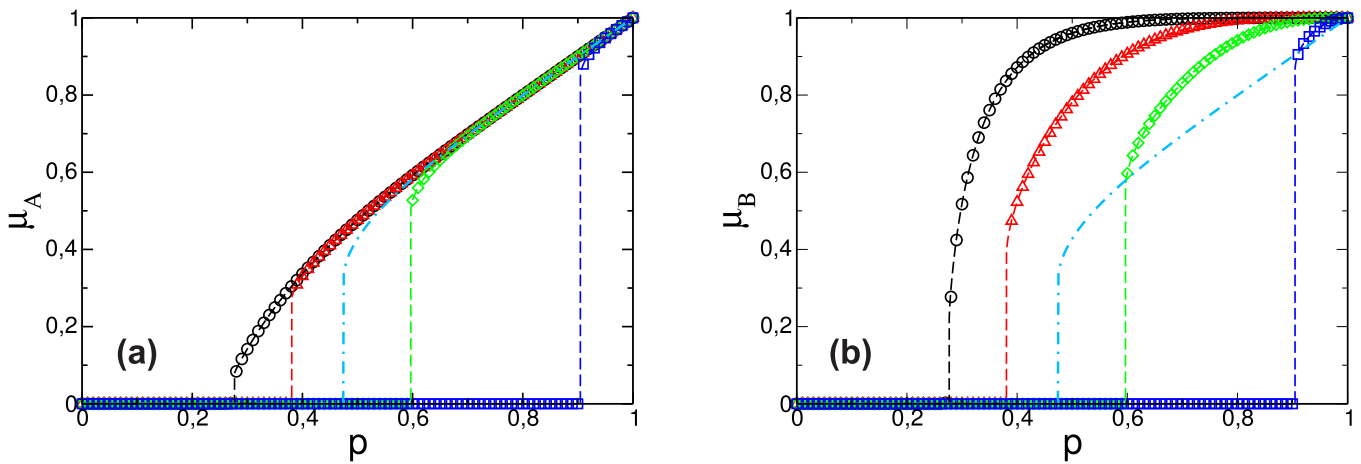}
\caption{Two random regular (RR) networks with
  $P_A(k)=P_B(k)=\delta_{k,5}$ and $P_{sA}(k)=P_{sB}(k)=\delta_{k,5}$
  and system size $N=10^5$ for different values of required supplies,
  $k_{sX}^*=1$ (\protect\unfbckcirc), $k_{sX}^*=2$
  (\protect\unfrdtrgle), $k_{sX}^*=3$ (\protect\ungrdiam),
  $k_{sX}^*=4$ (\protect\unfblubox), as a function of the initial
  fraction of survived nodes $p$. Also shown two RR networks with
  $P_A(k)=P_B(k)=\delta_{k,5}$ but $P_{sA}(k)=P_{sB}(k)=\delta_{k,1}$,
  $k_{sX}^*=1$ (\protect\skyline). The symbols are the results of the
  stochastic simulations and the lines are the iterated values
  obtained by equations (\ref{Eq.f} - \ref{Eq.mu}). The dashed-dotted
  lines represent only the theoretical results since they have been
  obtained in Ref[11]. In panels (a) and (b) we show the order
  parameter of network $A$ and $B$, $\mu_A$ and $\mu_B$, respectively
  for the giant component rule.}\label{PinfRR}
\end{figure}

Figure~\ref{PinfRR} shows a plot of $\mu_A$ and $\mu_B$ in the steady
state as a function of the initial fraction of surviving nodes $p$
when the giant component rule is applied.  The results for the k-core
rule are shown in the Supplementary Information. We use two random
regular (RR) networks with a degree distribution
$P_X(k)=\delta_{k,5}$, with $X=A,B$, and where the distribution of
supplies is also RR with $P_{s,A}(k)=P_{s,B}(k)=\delta_{k,5}$. For the
external rule of functionality we use $r_{sX}(j,k)=0$ if $j<m$ and
$r_{sX}(j,k)=1$ if $j\geq m$ for all $m$ from $m=1$ to $m=4$. The
results obtained from the equations (dashed lines) agree with the
results of the simulations (symbols). In addition we compare the
results of the present model with the results of the original model of
cascading failures \cite{Buldyrev2010} shown as a dashed-dotted line in which
$P_X(k)=\delta_{k,5}$, but $P_{s,A}(k)=P_{s,B}(k)=\delta_{k,1}$ and
$m=1$.

Note that in network $A$ the order parameter for all values of
$k_{s}^*$ is proportional to $p$ until it begins to drop and become
close to the critical threshold $p_c$. This means that the depletion
of the supply from network $B$ does not significantly impact network
$A$ until it reaches the collapse threshold at which the system breaks
down with a discontinuous transition. We calculate this critical value
numerically using the generating functions (see Supplementary
Information: section {\it Numerical solution for the threshold
  $p_c$}). Note also that, as expected, the behavior of network $B$ is
different. Because there is no initial random failure in network $B$,
it remains more intact than network $A$. When network $A$ crumbles,
however, both networks collapse. Thus despite its damage being minor
the transition in network $B$ is more abrupt, more unexpected, and,
therefore, more dangerous. This is the key difference between the
present mode and the original model \cite{Buldyrev2010} in which the
behaviors of network A and B are identical.  In addition, note that
the system is more resilient when $k_{s}^*$ is smaller, i.e., when the
supply level decreases. We also observe that the interdependent system
with only one supply-demand link (the dashed-dotted line) it is more
resilient than a system with more connections between the two
networks, but with large functionality thresholds $m \geq 3$.

If instead of the giant component we apply the k-core as an internal functionality rule
we get the same qualitative results. For different values of $k^*$ and
$k_{s}^*$ the order parameters also undergo a discontinuous transition,
and the system becomes more vulnerable when the threshold of internal
links and the threshold of supply links increases (see Supplementary
Information: section {\it k-core Percolation}).

\begin{figure}[ht] 
\centering
\includegraphics[width=\linewidth]{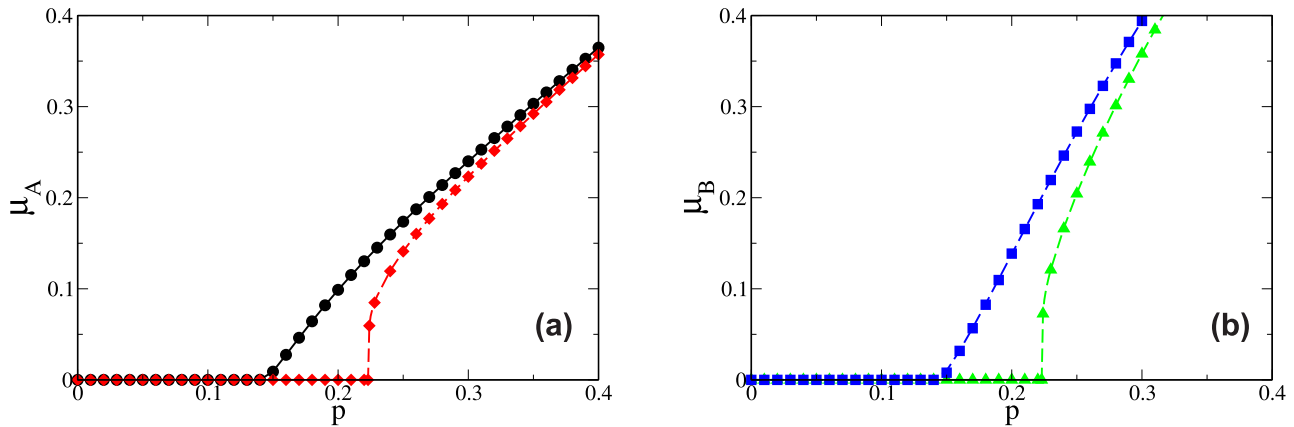}
\caption{Order parameters for the ``mass rule'', for a system of
  networks with internal distribution
  $P_{A}(k)=P_{B}(k)=\delta_{k,5}$, supply distributions
  $P_{sA}(k)=P_{sB}(k)=\delta_{k,2}$ and thresholds
  $k_A^*=k_B^*=1$. All the components of size $h=1$ are deleted
  ($q(1)=1$), and all the components of size $h\geq3$ are preserved
  ($q(3)=q(4)=...=q(h_{max})=0$ where $h_{max}$ is the maximum value
  of $h$). The curves represent the case $q(2)=0.3$ (\protect\bckcirc,
  \protect\blubox), for which there is a continuous transition, and
  $q(2)=0.85$ (\protect\rddiam, \protect\grtrgle), which leads to an
  abrupt breakdown of the order parameter. The dashed lines represent
  the theoretical results and the symbols the stochastic
  simulations. (a) Network $A$, (b) Network $B$.}\label{Finit}
\end{figure}

When applying the ``mass'' rule, finite components of size $h$ in
network $X$ survive with a probability $1-q_X(h)$. When all nodes have a
single supply-demand link, i.e., when $k_s=1$ and $k_s^*=1$, and all
finite components of size greater than or equal to $h=2$ are preserved,
the system undergoes a continuous transition \cite{DiMuroPRE}. Here
$q_X(1)=1$ and $q_X(h)=0$ for $h \geq 2$. If the number of supply links
increases and the threshold $k_s^*=1$ is fixed, the system becomes more
resilient and the transition remains continuous. In contrast, if all the
components of size $h=2$ are removed [$q_X(2)=1$] the transition becomes
discontinuous irrespective of the number of supply-demand links
connecting the networks. Nevertheless, not all the components of size
$h=2$ need to survive to have a continuous
transition. Figure~\ref{Finit} shows the order parameters for $q(2)=0.3$
and $q(2)=0.85$ when $q_A(h)=q_B(h)=q(h)$. Note that when $q(2)=0.3$ the
transition is continuous even when some of the components of size $h=2$
are deleted. When $q(2)=0.85$ the number of surviving $h=2$ components
is insufficient to prevent an abrupt transition.

\begin{figure}
  \includegraphics[width=350pt]{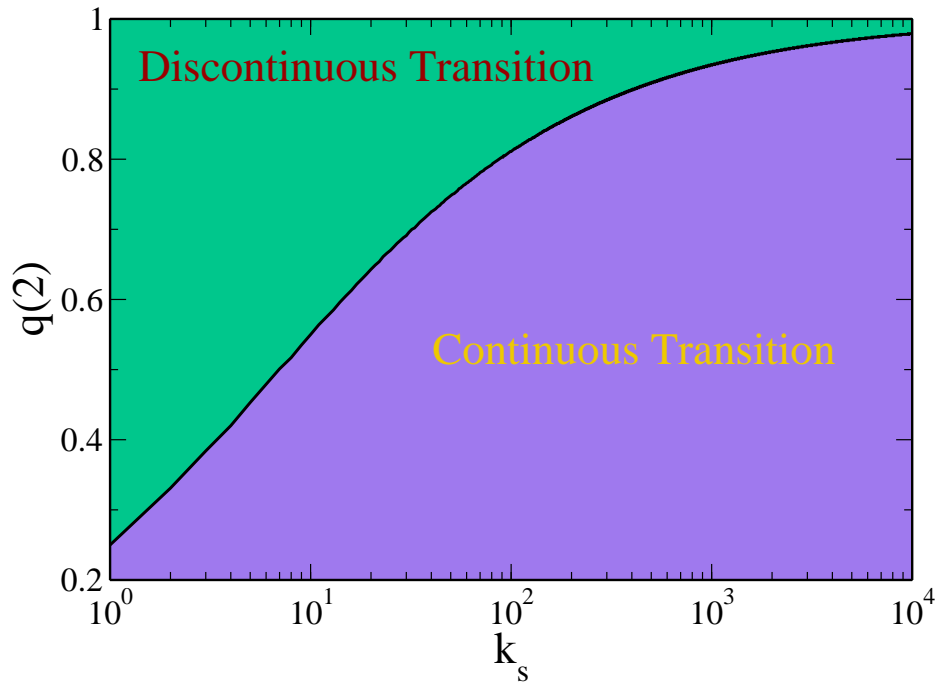}
\caption{Phase diagram that shows the continuous and discontinuous
  transitions zones when the ``mass rule'' is applied. The curve
  represents the critical probability of failure of the components of
  size $h=2$ as a function of the number of the supply-demand links. In
  this case $P_{A}(k)=P_{B}(k)=\delta_{k,5}$,
  $P_{sA}(k)=P_{sB}(k)=\delta_{k,k_s}$ and $k_{sA}^*=k_{sB}^*=1$. For
  clarity, the $k_s$ axis is shown on a log scale.}\label{Transition}
\end{figure}

\begin{figure}[ht] 
\centering
  \includegraphics[width=270pt]{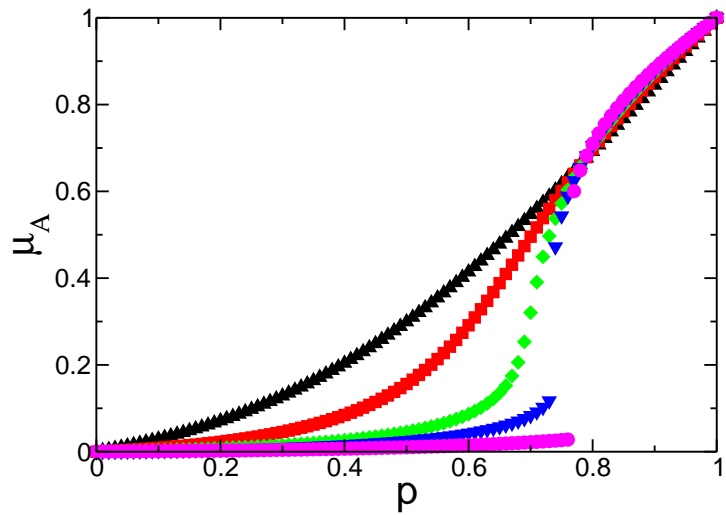}
\caption{Order parameter of network $A$ as a function of the initial
  failure for a bipartite system and for a threshold function
  $r_{sX}(j,k_{sX})=3(j/k_{sX})^2-2(j/k_{sX})^3$. The supply-demand
  distribution is single valued with $k_{sX}=3$ (\protect\bcktrgle),
  $k_{sX}=5$ (\protect\rdbox), $k_{sX}=7$ (\protect\grdiam), $k_{sX}=8$
  (\protect\invblutrgle) and $k_{sX}=10$ (\protect\mgcirc). For
  $k_s\geq8$ there is a discontinuous transition. The curves were
  obtained from the equations.  }  \label{Bipart}
\end{figure}

\begin{figure}[ht]
  \centering
\includegraphics[width=\linewidth]{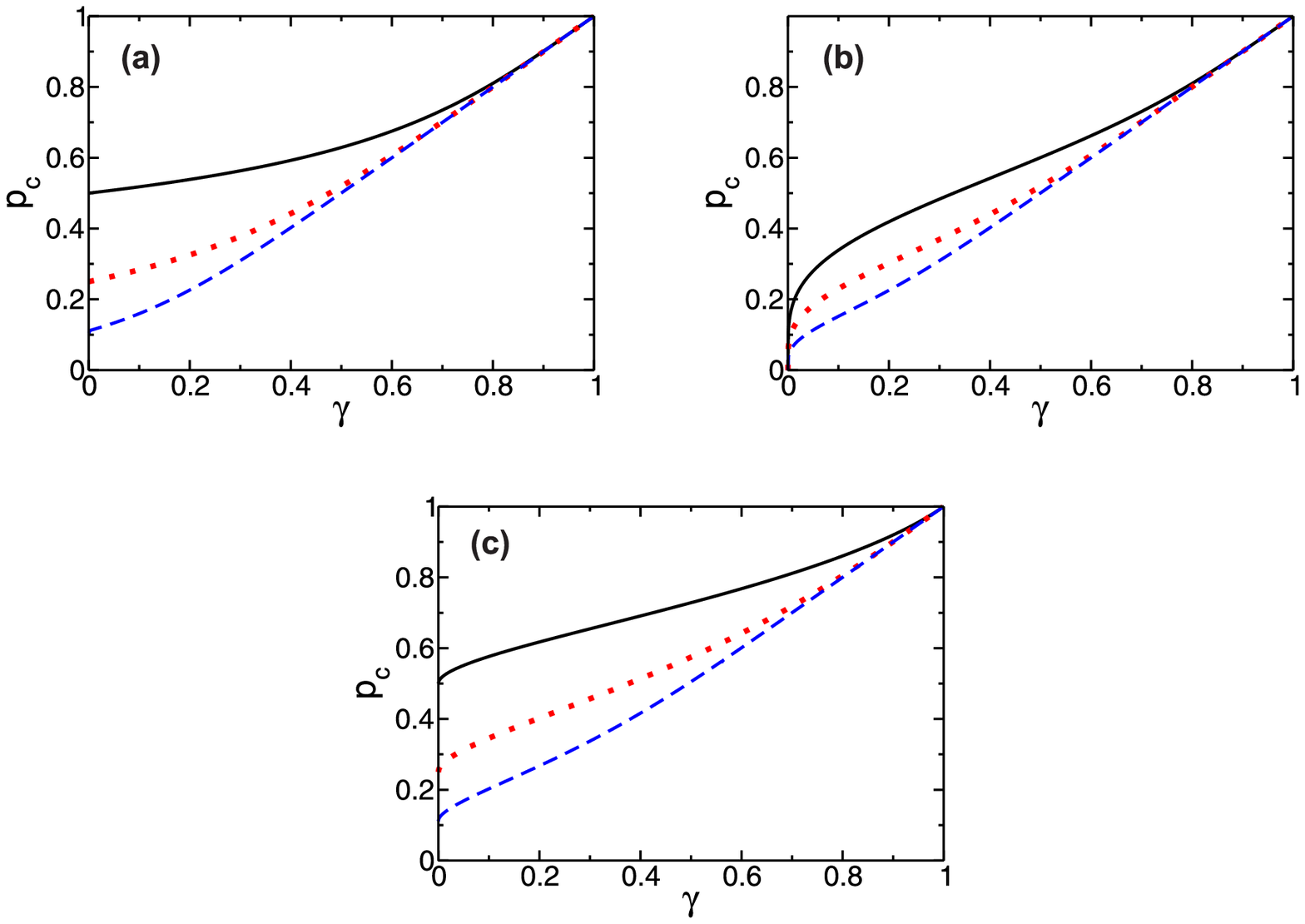}
\caption{Critical threshold $p_c$ as a function of
  $\gamma=k_{sB}^*/k_{sB}$ for different values of $z_A$, the internal
  connectivity of network $A$, where its internal degree distribution
  is RR. The curves represent different values of $z_A$: $z_A=3$
  (\protect\bckline), $z_A=5$ (\protect\rddotted) and $z_A=10$
  (\protect\bludash). Panel (a) corresponds to the Giant Component
  rule. Panel (b) corresponds to the ``mass rule'', with $q(h)=1$ for
  $h=1,2,3$, and panel (c) to the k-core rule with $k^*_X=2$. Note
  that in panel (b) $p_c \sim \gamma^{1/4}$ when $\gamma\to 0$, and
  thus corresponding curves appear finite even for very small
  $\gamma>0$.}\label{pcgamfig}
\end{figure}

Thus when $k_s^*=1$ there is a critical value of $q(2)=q_c(2)$ that
separates the zone of continuous transition from the zone of
discontinuous transition. Figure~\ref{Transition} shows a phase diagram
for a system of networks following the ``mass'' rule with an internal
distribution $P_{A}(k)=P_{B}(k)=\delta_{k,5}$ and supply distribution
$P_{sA}(k)=P_{sB}(k)=\delta_{k,k_s}$. Note that the behavior of the
critical probability as a function of the number of supply-links $k_s$
between the networks delimits these two zones. As $k_s$ increases the
system becomes more robust, and more components must fail to cause an
abrupt transition. In the limiting case $k_s\to \infty$ the curve
reaches the value $q_c(2)=1$, but also $p_c\to 0$. On the other hand,
when $k^*_s>1$ the transition is always discontinuous for any value of
$q(s)$ and sufficiently large $k_s$.

What happens if no internal functionality rule is applied? This could be
the case in a bipartite system in which nodes within each network do not
interact but use nodes in the other network as bridges to establish
connections. Here the exacerbation factor is simply $g_X(y)=1$, which
simplifies the equations. If we analyze this system for different
functions $r_{sX}(j,k)$ (see Supplementary Information: section {\it
  Examples of $r_{sX}(j,k)$ functions}) we see that if $r_{sX}(j,k)$ is
a step function with fixed threshold $k_{sX}^\ast=2$, the transition is
continuous, but it is discontinuous for $k_{sX}^\ast>2$, and there is no
transition for $p>0$ if $k_{sX}^\ast=1$. Also if we choose a linear
function, i.e., $r_{sX}(j,k_{sX})=j/k_{sX}$, there is again no
transition because here functions $W_s(\beta)$ and $Z_s(\beta)$ become
linear functions of $\beta$. On the other hand, when the function $r_{sX}$ is
nonlinear, the behavior changes. Figure~\ref{Bipart} shows the behavior
of the order parameter of network $A$ for a polynomial function
$r_{sX}(j,k_{sX})=3(j/k_{sX})^2-2(j/k_{sX})^3$ and for a supply-demand
distribution $P_{s,X}(k)=\delta_{k,k_s}$. Note that for small values of
$k_s$ the order parameter moves smoothly to zero but for $k_s=8$ the
system undergoes a discontinuous transition. The existence of these
transitions can be explained studying Eqs.~(\ref{Eq.f}) and (\ref{Eq.y})
(see Supplementary Information: section {\it Numerical solution for the
  threshold $p_c$}).

Unlike the previous results, the transition here does not produce a
total collapse of the system, and after the jump a small fraction of
nodes remains functional for any $p>0$. If a delta-distribution of
supply links is replaced by the Poisson distribution with $\langle
k_{s}\rangle_X=\lambda$, we find a critical point on a
$(p,\lambda)$ plane $\lambda_c=7.58465, p_c=0.728102$ at which the
first order phase transition emerges. For $\lambda>\lambda_c$ the
transition is first order and for $\lambda<\lambda_c$ there is no
phase transition for $p>0$. At this point the system belongs to the
mean-field universality class, such as the Ising model in infinite
dimensions where $p$ corresponds to the ordering field and $\lambda$
to the thermal field.

We next analyze the limiting case of large $k_s$ values when all nodes
in network $B$ have a fixed threshold $k^*_{sB}$, and we find that the
critical point $p_c$ converges to a value determined by the ratio
$\gamma\equiv k_{sB}^*/k_{sB}$ given by
\begin{eqnarray}\label{Eq.pcg}
\gamma=p_c g_A(p_c),
\end{eqnarray}
which is valid for all of the internal failure rules. 

The $p_c$ value depends on $\gamma$ in this limit because when $\langle
k_{sX}\rangle \to \infty$ the functions $W_{sB}(\beta)$ and
$Z_{sB}(\beta)$ become step functions equal to $0$ for $\beta<\gamma$
and to $1$, otherwise.
Note that $\gamma$ only relates to the external
properties of network $B$, but that the value of $p_c$ depends solely on
the topology of network $A$. This is because network $B$ is intact above
$p_c$, but when $p<p_c$ all the supply-demand links maintaining the
integrity of network $B$ fail and the entire structure crumbles. Thus
here the topology of network $B$ does not affect the final state of the
system. See Supplementary Information: section {\it Asymptotic
  properties of the functions $W_s$ and $Z_s$} for the derivation of
Eq.~(\ref{Eq.pcg}).

Figure~\ref{pcgamfig} shows the behavior of Eq.~(\ref{Eq.pcg}) for each
internal rule of functionality and for several values of internal
connectivity $z_A$ in network $A$ when it has an internal degree
distribution $P_A(k)=\delta_{k,z_A}$. Note that all curves go to
$p_c=1$ when $\gamma \to 1$, i.e., $k_{sB}^* \sim k_{sB}$, and thus
even a small perturbation can cause a system breakdown. In contrast,
curves with higher $z_A$ values have lower $p_c$ values because
increased connectivity means increased resilience. In addition, when
$\gamma \to 0$ then $k_{sB} \gg k_{sB}^*$, rendering the influence of
network $B$ on network $A$ insignificant. Here network $A$ behaves as
an isolated system. We see this in the giant component rule [see
  Fig.~\ref{pcgamfig}(a)] in which $p_c \to 1/(z_A-1)$ as $\gamma \to 0$,
a value that corresponds to the critical threshold of node percolation
\cite{Cal_01,New_01} in isolated RR networks. Similarly, for the
``mass'' rule we find that the threshold behaves as $p_c\to 0$ when
$\gamma \to 0$ because when there is an initial attack $1-p$ on an
isolated network there are always components of varying masses in the
thermodynamic limit (with an infinite number of nodes). Thus when
$q(h)<1$ for any size $h$ there are always surviving components when
$p>0$.

If there is a Poisson internal degree distribution in network $A$, i.e.,
$P_A(k)=exp\big[-\langle k \rangle_A\big]\langle k \rangle_A^k/k!$
where $\langle k \rangle_A$ is the mean connectivity, we can write a
closed-form expression for $p_c$ for the giant component rule,
\begin{equation}\label{ERpcmain}
p_c=\frac{\gamma}{1-exp\big[-\gamma\;\langle k \rangle_A  \big]}.
\end{equation}

Note that $p_c$ does not depend on the internal degree distribution of
network $B$. The derivation of Eq~(\ref{ERpcmain}) is supplied in the
Supplementary Information: section {\it Asymptotic properties of the
  functions $W_s$ and $Z_s$}. On the other hand, if the system is
bipartite then from Eq.~(\ref{Eq.pcg}) the critical value is simply
$p_c=\gamma$.

\section*{Discussion}

\noindent
We have analyzed the cascading failure process in a system of two
interdependent networks in which nodes within each network have multiple
connections, or supply-demand links, with nodes from their counterpart
network. In this model each node must have at least a given number of
supply-links leading to functional nodes in the other network to remain
active. We call this number the supply threshold and we call this
condition the external functionality rule.  We have studied the process
under three internal functionality rules, (I) nodes must belong to the
giant component in their own network, (II) nodes that belong to a finite
component survive with a probability determined by the mass of the
component, and (III) an internal version of the external functionality
rule, known as heterogeneous k-core percolation. In addition, we have
studied a system in the absence of any internal functionality rule,
which is equivalent to a bipartite network. Our system is a
generalization of the models of interdependent networks
\cite{Buldyrev2010,Parshani} that represent a particular case of our
model with $P_{sX}(k)=0$ for $k>1$ and a giant component rule of
internal functionality. Our model shows a rich behavior for various
parameter values that is characterized by the appearance of
discontinuous first order transitions. In some cases, multiple first
order transitions can be observed, a situation impossible in the original
models\cite{Buldyrev2010,Parshani}.

We have found that for all the internal functionality rules the system
is more robust when the supply threshold is lower. Under internal
rules I and III there is a discontinuous transition at a collapse
threshold $p=p_c$. The main difference between our model and the
previously studied models \cite{Buldyrev2010,Parshani} is that in the
case of multiple supply links the initial attack on network $A$ does
not immediately affect network $B$, and it remains more functional
than network $A$ for any $p>p_c$.  This makes the transition, when it
occurs in network $B$, more abrupt than in network $A$. These sudden
breakdowns can come without warning. In some catastrophic events,
e.g., an earthquake of sub-threshold strength, the damage to network
$B$ may be minor and the development of precautions or recovery
strategies thus deemed of minor importance. This becomes problematic
when the strength of an earthquake exceeds a certain threshold and
causes a total breakdown in network $B$.  In contrast, in ``mass''
rule II for $k_s^*=1$ the transition can be continuous depending on
the probability that components of size $h=2$ remain functional and on
the number of supply-demand links. For each value of $k_s$ there is a
critical probability $q(2)$ below which the transition becomes
discontinuous.

When the model is applied to a bipartite system, the behavior is
determined by function $r_{sX}$. In particular, when this function is
polynomial there is no transition in $k_{sX}\leq7$, but when $k_s$
increases this curve breaks and becomes discontinuous.

Finally we have studied the asymptotic limit value of the number of
supply-demand links, and find that when $r_{sB}$ is a step function
there is an exact relationship between the ratio $\gamma =
k^\ast_{sB}/k_{sB}$ and the collapse threshold $p_c$. We also find
that in this limit the resilience of the interacting system is
enhanced up to the point at which the critical threshold $p_c$ is
solely dependent on the topology of network $A$.

\section*{Methods}\label{Sec-meth}

\noindent
For the stochastic simulations we use for both networks a system size of
$N=10^6$ to compute the steady state and $N=10^8$ for the temporal
evolution close to the critical threshold (See Fig.~\ref{Pinf_n}). We
use the Molloy-Reed Algorithm \cite{Mol_01} for the construction of the
networks. The simulation results are averaged over $1000$ network
realizations.

For model II, the ``mass'' rule, a finite component of size $h$ survives
with probability $1-q(h)$. In the stochastic simulations if a finite
component remains after the internal failure at a step of the cascade,
then in the following steps of the cascade this component only can fail
due to the external rule of functionality.

In our theoretical analysis, to calculate the values of the order
parameters at the steady state we iterate the temporal evolution
Eqs.~(\ref{Eq.f})--(\ref{Eq.mu}) until the condition $\mu_A
\equiv\mu_{A,n} =\mu_{A,n-1}$ is satisfied. At this stage the magnitudes
of all order parameters reach a steady state and no longer change.

\section*{{\Large Supplemental Information}}

\section{Explicit form of the functionality rules}\label{S.explicit_supp}

\subsection{Giant component} \label{SecIA}
The giant component in a network is the largest connected
component. Most functioning networks are completely connected, but when
they experience failure, finite components---little islands of
nodes---become disconnected from the giant component. A common
functionality rule states that nodes in these finite components have
insufficient support to remain active. Thus in addition to the nodes
rendered inactive by failure, the exacerbation factor renders inactive
all nodes not connected to the giant component. If network $X$ has a
degree distribution $P_X(k)$ and a fraction $1-y_X$ of nodes is randomly
removed, the exacerbation factor $g_X$ is
$g_X(y_X)=1-G_0^X[1-y_X(1-f_\infty^X)]$, where $f_\infty^X$ is the
probability that the branches do not expand to infinity, and it
satisfies the recurrent equation
$f_\infty^X=G_1^X[1-y_X(1-f_\infty^X)]$. The functions $G_0^X(u)$ and
$G_1^X(u)$ are the generating functions of the degree distribution and
the excess degree distribution, respectively. They are given by
$G_0^X(u)=\sum_k P_X(k)u^k$ and $G^X_1(u)=\sum_k k/\langle k \rangle_X
P_X(k)u^{k-1}$, where $\langle k \rangle_X$ is the average connectivity
of network $X$, $\langle k\rangle_X=\sum_k k\;P_X(k)$ .

\subsection{Finite components} 
We can relax the giant component rule and allow some finite components
to be self-sustaining and remain functional. If we allow the giant
component to remain active after a failure and also some of the finite
components to remain active with a probability related with their size
$h$, then the exacerbation factor is
\begin{equation}
g_{X}(y_{X})=1-\sum_h q_X(h)\pi_{h,X}(y_X), 
\end{equation}
where $q_X(h)$ the probability that a component of size $h$ has been
removed, and $\pi_{h,X}(p)$ the probability that a randomly-selected
surviving node belongs to a component of size $h$. We can obtain the
functions $\pi_{h,X}(y_X)$ using the Lagrange inversion formula
\cite{DiMuroPRE} for any given distribution $P_X(k)$.

\subsection{k-core Percolation}
In conventional or homogeneous k-core percolation, every node has an
identical threshold $k^\ast$.  Thus following a failure, if the number
of surviving nodes among the $k$ neighbors of a node is less than
$k^\ast$, the node fails, otherwise it remains functional. In contrast,
in heterogeneous k-core percolation each node $i$ with initial degree
$k_i$ has a randomly assigned threshold $k^\ast_i\leq k_i$.  In
heterogeneous k-core percolation, the distribution of thresholds
$k^\ast_i$ is given by the cumulative distribution $r_{X}(j,k)=P(k^*
\leq j|k)$, where $k$ denotes the degree values of network nodes.  The
simplest $r_X(j,k)$ case is a step function, i.e., $r_{X}(j,k)=0$ if
$j<k^*$ and $r_{X}(j,k)=1$ if $j\geq k^*$ and for all $k$. This is
equivalent to assigning all nodes the threshold $k^\ast_i=k^\ast$, which
is equivalent to homogeneous k-core percolation. Another option is the
linear function $r_{X}(j,k)=k^*/k$ in which the thresholds $k^\ast$ for
nodes with an initial degree $k$ are uniformly distributed between $1$
and $k$.

If we know the degree distribution $P_X(k)$ and the threshold
distribution $r(j,k)$, we can define the heterogeneous k-core generating
function
\vspace{0.5cm}
\begin{equation}
W_X(\beta)=\sum_{k=0}^\infty P_X(k)\sum_{j=0}^{k} \binom{k}{j}
r_{X}(j,k)\beta^{j}(1-\beta)^{k-j}, 
\end{equation}
and the $k$-core generating function of the excess distribution,
\begin{equation}
Z_X(\beta)=\sum_{k=1}^\infty \frac{k P_X(k)}{\langle
  k\rangle_X}\sum_{j=0}^{k-1} \binom{k-1}{j}
r_{X}(j+1,k)\beta^{j}(1-\beta)^{k-j-1}.
\end{equation}

The exacerbation factor of the k-core heterogeneous percolation can thus
be written $g_X(y_X)=W_X(\beta)$, where as in Sec.~\ref{SecIA} $\beta$
satisfies the self-consistent equation $\beta=y_X \: Z_X(\beta)$, and
$y_X$ is the fraction of surviving nodes in network $X$.

\begin{figure}[H]
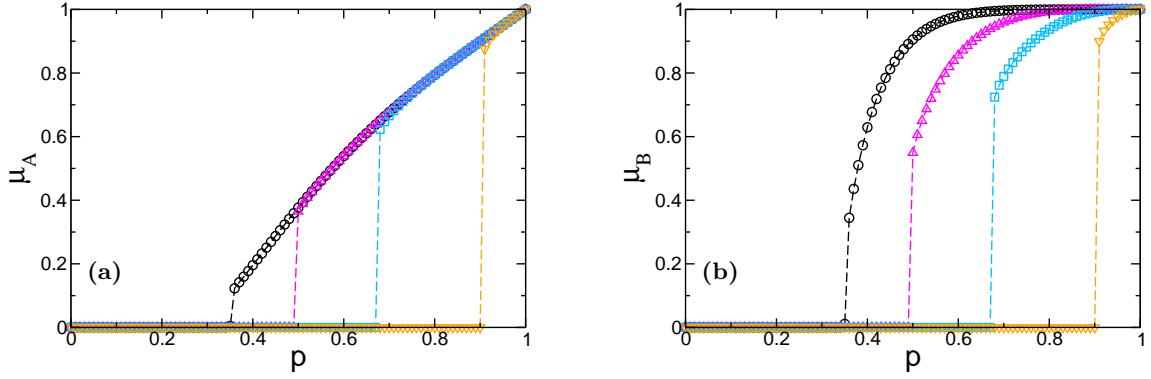
 
\vspace{1cm}
\begin{center}
  \begin{overpic}[scale=0.28]{m2_redA.eps}
    \put(15,18){\bf{(a)}}
  \end{overpic}\vspace{1cm}
  \hspace{1cm}
   \begin{overpic}[scale=0.28]{m2_redB.eps}
    \put(15,18){{\bf{(b)}}}
  \end{overpic}\vspace{1cm} 
\end{center}
\vspace{-2cm} \caption{Order parameters for the homogeneous k-core rule
  with $k^\ast=2$ as a function of the initial fraction of survived
  nodes $p$, for two random regular (RR) networks with
  $P_{sA}(k)=P_{sB}(k)=\delta_{k,5}$ and system size $N=10^5$, and
  different values of required supplies, $k_{s}^*=1$
  (\protect\unfbckcirc), $k_{s}^*=2$ (\protect\unfmgtrgle), $k_{s}^*=3$
  (\protect\unfcybox) $k_{s}^*=4$ (\protect\unfinvogtrgle). The symbols
  are the results of the stochastic simulations and the lines the
  iterated values from the equations. (a) Network $A$. (b) Network
  $B$.}\label{PinfRRkcor}
\end{figure}

Figure~\ref{PinfRRkcor} plots $\mu_A$ and $\mu_B$ in the steady state
for the k-core rule.

\section{Numerical Solution for the threshold $p_c$}\label{S.0}
The critical point $p_c$ at which the transition takes place can be
determined using the equations from the main text.  We combine the
equations in set (3) at the steady state at which
$f_{X,n}=f_{X,n-1}=f_{X}$, thus withdraw the $f_{B}$-dependence, and
obtain an equation in terms of $f_{A}$,
\begin{equation}\label{fabself} 
f_A=F(f_A)\equiv Z_{sB} \big[\eta(f_A) \big]  g_B\big[ W_{sB}[\eta(f_A)
  ]  \big], 
\end{equation}
with $\eta(f_A)\equiv f_B=p Z_{sA}\big(f_A\big)g_A\big[p W_{sA}(f_A)\big]$.

\begin{figure}[h]
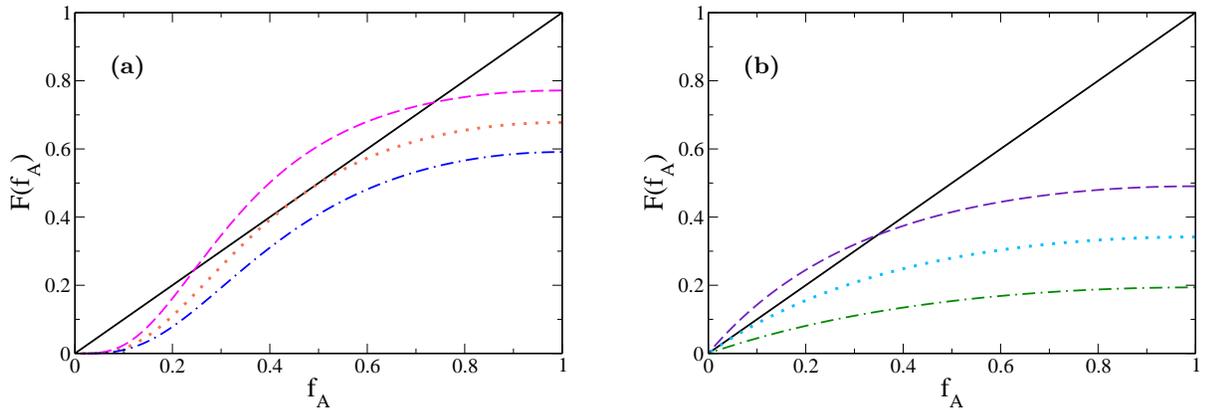
 
\vspace{1cm}
\begin{center}
  \begin{overpic}[scale=0.3]{Discontinuous.eps}
    \put(18,60){\bf{(a)}}
\end{overpic}\hspace{1cm}\vspace{1cm}
  \begin{overpic}[scale=0.3]{Continuous.eps}
    \put(18,60){{\bf{(b)}}}
 \end{overpic}\hspace{1cm}\vspace{1cm}
\end{center}
\vspace{-2cm} \caption{Graphical solution of Eq.~(\ref{fabself}) for a
  system of two RR networks with $z=5$, $P_{sX}(k)=\delta_{k,2}$, in
  which all nodes have the same threshold $k_s^*=1$. In this case we use
  the ``finite components'' rule, with $q_X(1)=1$ and $q_X(h)=0$ for
  $h>2$. In (a) we have $q_X(2)=1$, which results in a discontinuous
  transition. The curves represent different values of $p$: $p=0.26$
  (\protect\magline), $p=0.22$ (\protect\bluline) and $p=p_c=0.2374$
  (\protect\orgline). For the critical threshold, the curve is tangent
  to the identity at the solution $f_A>0$. In (b) we show a continuous
  transition with $q_X(2)=0$, in which $p=0.13$ (\protect\indigline),
  $p=0.07$ (\protect\verdline) and $p=p_c=0.1$ (\protect\skylinex). For
  the curve that represents the critical value, the point of tangency is
  located at $f_A=0$.}\label{graphsol}
\end{figure}

For a given value of $p$, the solution is the intersection between
function $F$ and the identity. Above $p_c$ there is always a non-trivial
solution. When $p<p_c$, Eq.~(\ref{fabself}) is only valid for
$f_A=0$. The method of finding the critical point differs depending on
whether the transition is discontinuous or
continuous. Figure~\ref{graphsol} shows the graphical solution of
Eq.~(\ref{fabself}) for (a) a discontinuous transition and (b) a
continuous transition. We plot the curves for $p=p_c$, $p>p_c$, and
$p<p_c$. When the transition is abrupt, when $p=p_c$ function $F$ is
tangent to the identity at $f_A=f_{Ac}$, which is the solution to
Eq.~(\ref{fabself}) for the critical threshold. Thus we have a condition
that must be fulfilled at the critical point,
\begin{equation}\label{FDERIV}
  \frac{dF(f_A)}{df_A}=1.
 \end{equation}
Thus we can solve Eqs.~(\ref{fabself}) and (\ref{FDERIV}) numerically to
find the critical threshold $p_c$ for a discontinuous abrupt transition.

In contrast, Fig.~\ref{graphsol}(b) shows that function $F$ is also
tangent to the identity, but here at $f_A=0$. Thus for a continuous
transition we can find the critical value $p_c$ by solving
Eq.~(\ref{FDERIV}) when $f_A=0$.

For the mass rule, the type of transition is related to the survival
probability of the $h=2$ components. If we assume here that all the
$h=1$ components and a fraction $q(2)$ of the $h=2$ components fail,
then for single-value internal degree distributions and supply
distributions when $k^*_s=1$, the derivative of function $F$ at the
critical threshold evaluated at the origin is
\begin{equation}\label{FD0}
\frac{d F(f_A)}{d f_A}
\bigg\rvert_{f_A=0}=\Big(p_c\;k_s\;k\;[1-q(2)]\Big)^2. 
\end{equation}
Thus when $q(2)=1$ the continuity condition is fulfilled only for $k_s
\to \infty$, as we can see in Fig.~5 in the main text.

On the other hand we can use Eq.~(\ref{fabself}) to understand the
transitions present in Fig.~6 from the main text. Here the system is
bipartite and thus we apply no internal functionality rule. The system
also has a supply distribution $P_s(k)=\delta_{k,k_s}$ and a supply
threshold distribution $r_{sX}(j,k)=3(j/k)^2-2(j/k)^3$.

\begin{figure}[h]
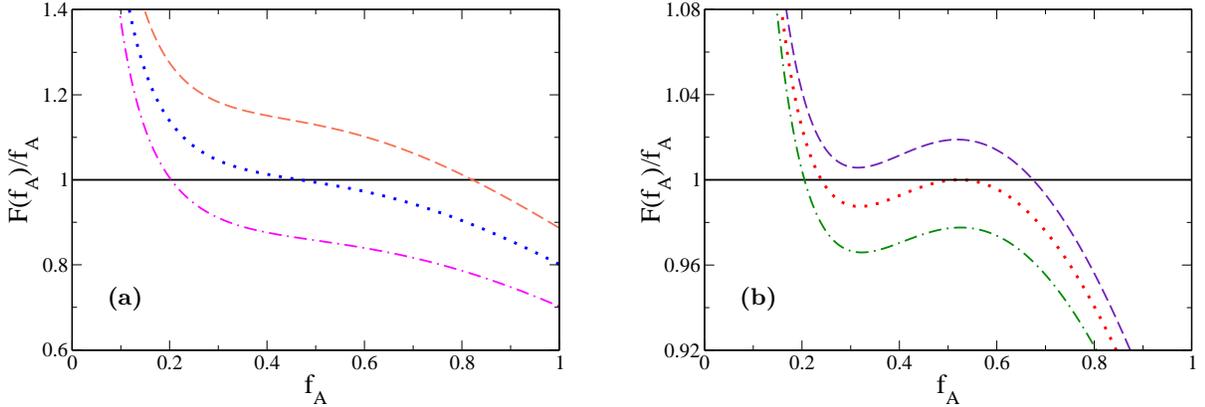
 
\vspace{1cm}
\begin{center}
  \begin{overpic}[scale=0.3]{Bipart_ks_7.eps}
    \put(18,18){\bf{(a)}}
\end{overpic}\hspace{1cm}\vspace{1cm}
  \begin{overpic}[scale=0.3]{Bipart_ks_8.eps}
    \put(18,18){{\bf{(b)}}}
 \end{overpic}\hspace{1cm}\vspace{1cm}
\end{center}
\vspace{-2cm} \caption{Graphical solution of Eq.~(\ref{fabself}) for a
  bipartite system with a supply distribution $P_s(k)=\delta_{k,k_s}$,
  and a supply threshold distribution
  $r_{sX}(j,k)=3(j/k)^2-2(j/k)^3$. Unlike Fig.~\ref{graphsol} here the
  $y$ axis is divided by the $x$ axis to have a better visualization of
  the intersection points. In (a) we have $k_{sX}=7$, value for which
  there is a continuous transition for $p=0$, as for $p>0$ the curve
  always intersect the identity at a non-zero point. The curves
  represent $p=0.6$ (\protect\magdashdotline), $p=0.7$
  (\protect\bludotline) and $p=0.8$ (\protect\orgdashline) . On the
  contrary in (b) we have $k_{sX}=8$ and there is a discontinuous
  transition. For $p=0.75$ (\protect\indigline) there is only one point
  of intersection, but for lower values of $p$ two more solutions
  appear, although the largest of them is the one related with the
  process. For $p=0.73626$ (\protect\reddotline) two of the solutions
  fuse together, and the curve becomes tangent to the identity, thus the
  solution for this value of $p$ is the point of tangency. However for
  lower values of $p$ the point of intersection of the left is the only
  solution of the process, as we can observe for $p=0.72$.
  (\protect\verdline)} \label{graphsolbipart}
\end{figure}

Figure~\ref{graphsolbipart} shows the graphical solution for this system
for different values of initial failure
$p$. Figure~\ref{graphsolbipart}(a) shows $k_s=7$. Note that the curve
always intersects the identity at a non-zero value when $p>0$. Thus the
order parameter goes to zero in a continuous transition when $p=0$. In
contrast, Fig.~\ref{graphsolbipart}(b) shows $k_s=8$, and the curve
behaves differently. For large values of $p$ there is only one solution,
but decreasing $p$ produces a new lower stable solution. A point of
intersection is a stable solution when a iterative process converges to
this point. If initially $f_A=1$, i.e., network $B$ is intact at the
beginning of the cascade\footnote{Recall that f A is the probability of randomly selecting a supply-demand link that leads to a functional
node in network $B$}, then successive iterations of
Eq.~(\ref{fabself}) converge to the highest solution. However when $p$
further decreases there is a critical value at which the largest
solution suddenly vanishes, and the iterative process converges to the
lower point, which is now the only solution. This abrupt change of
solutions causes the discontinuous transition shown in Fig.~6 in the
main text.

\section{Asymptotic properties of the functions $W_s$ and $Z_s$}\label{S.1}
We next study the case in which $r_{sX}$ is a step function
$\Theta(k_s-k_{s}^{*})$, i.e., when all nodes have the same threshold
$k_s^*$. Using the supply distribution $P_{sX}(k)=\delta_{k,k_s}$ for
simplicity, we show that the behavior of the order parameters does not
depend on the values of $k_s^*$ and $k_s$ when $k_s$ is large, but on
the ratio $k_s^*/k_s \equiv \gamma$. We assume that networks $A$ and $B$
have the same external properties, and we drop index $X$.  Here we
rewrite function $W_s(x)$ [see Eq.~(1) from the main text] using the
incomplete beta function
\begin{eqnarray}\label{eq.IncompBet}
W_s(x)=1-(k_s-k_s^*+1)\binom{k_s}{k_s^*-1}
\int_0^{1-x}t^{k_s-k_s^*}(1-t)^{k_s^*-1}\;dt, 
\end{eqnarray}
and it is thus straightforward to compute the first derivative of $W_s$ 
\begin{eqnarray}\label{eq.IncompBet1}
W_s^{'}(x)=(k_s-k_s^*+1)\binom{k_s}{k_s^*-1}x^{k_s^*-1}(1-x)^{k_s-k_s^*}. 
\end{eqnarray}.

When $W_s$ converges to a step function for large $k_s$ its derivative
$W_s^{'}$ converges to a Dirac delta centered on $x_c$. To verify this
we calculate the mean of this function and its variance,
\begin{eqnarray}
  \langle x \rangle &=&\frac{k_s^*}{k_s+1}\\[10pt]\nonumber
  \sigma_x &=&\frac{k_s^*(k_s^*+1)k_s!}{(k_s+2)!} - \left(\frac{k_s^*}{k_s+1}\right)^2.
\end{eqnarray}

For a fixed value $\gamma \equiv k_s^*/k_s$ the variance goes to zero
when $k_s \to \infty$, indicating that function $W_s$ is discontinuous
at $x=x_c$. The previous analysis is also valid for $Z_s$, but when
$k_s^*>1$, since $Z_s(x)=1$ when $k_s^*=1$.

Thus $W_s(x)$ and $Z_s(x)$ (when $k_s^*>1$) converge to a Heaviside
distribution, which depends on the ratio $\gamma$ for $k_s \to \infty$,
\begin{equation}\label{Step}
W_s(x)=Z_s(x)=
\begin{cases} 
      0 & x< \gamma ; \\
      \frac{1}{2} & x=\gamma , \\
      1 & x> \gamma .
   \end{cases}.
\end{equation}
Then in this limit the solution of Eqs.~(1)--(3) in the steady state
and the transition point $p_c$ is dependent only on $\gamma$.

If we know the asymptotic properties of these functions, we can
determine the critical point $p_c$ behavior at this limit. As $p \to
p_c$ we expect that $f_B<f_A$, since network $A$ receives the initial
failure, i.e., the probability $f_B$ that an external link from network
$A$ to $B$ leads to a functional node in network $A$ is lower than in
the opposite direction. Then using Eq.~(\ref{Step}) at criticality
$Z(f_B)=W(f_B)=0.5$ and $Z(f_A)=W(f_A)=1$, i.e., $f_A>\gamma$
and \footnote{Note that when $Z_s(f_B)=0$ the interdependent network collapses, but when $Z_s(f_B)=1$ the functional
network is in the steady state.} $f_B=\gamma$. Then the set (1) equations in the main text
in the steady state at $p=p_c$ can be rewritten
\begin{eqnarray}
  f_A&=&\frac{1}{2}\;g_{B}\left(\frac{1}{2}\right),\\
  f_B&=&\gamma=p_c Z(f_A) g_A\big[p_cW(f_A)\big].\label{eq.fb_asym}
\end{eqnarray}
Because $f_A>\gamma$, $Z(f_A)=1$ according to Eq.~(\ref{Step}). Thus
from Eq.~(S12) 
\begin{eqnarray} \label{pcgam}
  \gamma=p_c g_A(p_c),
\end{eqnarray}
which is related to the value of $p_c$ with $\gamma$.

Note that when $k_s^*=1$, Eq.~(S13) becomes
$f_A=g_{B}\left(\frac{1}{2}\right)$. Nevertheless here Eq.~(\ref{pcgam})
still holds.

Using the giant component rule at this limit we find analytically the
value of the criticality threshold for a particular case. If network $A$
has an internal Poisson degree distribution, i.e., if $P_A(k)=\langle
k\rangle_A^k exp[-\langle k\rangle_A]/k!$ where $\langle k\rangle_A$ is
the average internal connectivity of network $A$, then
$G^A_0(x)=G^A_1(x)=exp[\langle k\rangle_A (x-1)]$. Here
$\gamma=p_c(1-f^A_\infty)$, and thus we obtain
\begin{equation}
  p_c=\frac{\gamma}{1-exp[-\gamma\;\langle k\rangle_A ]}.
\end{equation}

\section{Examples of $r_s(j,k)$ functions}

Equations
\begin{equation}\label{WWW}
W_{sX}(f)=\sum_{k=0}^\infty P_{sX}(k)\sum_{j=0}^k
r_{sX}(j,k)C_{k,j}f^j(1-f)^{k-j}. 
\end{equation}   
and 
\begin{equation}\label{ZZZ}
Z_{sX}(f)=\sum_{k=0}^\infty \frac{k P_{sX}(k)}{\langle k_s \rangle_X
}\sum_{j=0}^{k-1} r_{sX}(j+1,k)C_{k-1,j}f^{j}(1-f)^{k-j-1}, 
\end{equation}   
can be evaluated explicitly for the power law shape
\begin{equation}
r_{sX}(j,k)=\left(\frac{j}{k}\right)^m,
\end{equation}
when $k>0$ and $r_{sX}(0,0)=1$. The latter condition is to prevent
autonomous nodes with no supply links from dying. Obviously this shape
can be generalized to any polynomial by which any function $r_{sX}(j,k)$
can be approximated.  Successively applying operator $f d/df$ to the
corresponding probability generating function, i.e., binomials
$(f+q)^k$, and letting $q=1-f$ allows us to express functions
$W_{sX}(f)$ and $Z_{sX}(f)$ as polynomials of power $m$ of $f$
\begin{equation}
W_{sX}(f)=\sum_{n=0}^{m} w_n^{m}f^n
+P_{sX}(0)~~~~~~Z_{sX}(f)=\sum_{n=0}^m z_n^mf^n,
\end{equation}
with coefficients that can be expressed through negative moments of the
distribution $P_{sX}(k)$
\begin{equation}
w_n^m=S(m,n)\sum_{k=1}^\infty P_{sX}(k)\frac{k!}{k^m(k-n)!} 
\end{equation}
and
\begin{equation}
z_n^m=\frac{S(m+1,n+1)}{\langle k_s\rangle_{X}}\sum_{k=1}^\infty
P_{sX}(k)\frac{k!}{k^m(k-1-n)!},  
\end{equation}
where $S(m,n)$ are Stirling numbers of the second kind that obey
recursion relation $S(m+1,n)=nS(m,n)+S(m,n-1)$ with initial conditions
$S(0,0)=1$, $S(0,n)$=0 when $n>0$, and $S(m,0)=0$ when $m>0$.

For linear $r_{sX}$ ($m=1$), functions $W_{sX}(f)$ and $Z_{sX}(f)$ are
linear functions of $f$
\begin{equation}
W_{sX}(f)=P_{sX}(0)+\Big(1-P_{sX}(0)\Big)f
~~~~~~~~Z_{sX}(f)=\left(1-\frac{1-P_{sX}(0)}{\langle
  k_s\rangle_{X}}\right) f +\frac{1-P_{sX}(0)}{\langle k_s\rangle_{X}}. 
\end{equation}

In the case described in Ref.~\cite{Buldyrev2010},
$P_{sX}(k)=\delta_{1k}$ produces $W_{sX}(f)=f$ and $Z_{sX}(f)=1$, and
hence recursive Eqs.~(4) and (5) from the main text reduce to Eq.~(1) in
Ref.~\cite{Buldyrev2010}. Even when $P_{sX}(k)\neq\delta_{1k}$, linear
$r_{sX}$ leads to the same phenomenon described in
Ref.~\cite{Buldyrev2010}, i.e., there is a first order phase transition
when $P_{sX}(0)$, the fraction of the autonomous nodes is small, and
there is a second order phase transition when $P_{sX}(0)$ is large.

On the other hand, there are differences in a bipartite system of two
networks in which the exacerbation factor is $g_X(y)=1$. Here linear
$r_{sX}$ does not produce a non-trivial phase transition when $p>0$
because Eqs.~(\ref{WWW}) and (\ref{ZZZ}) become linear equations of $f$,
but threshold $r_{sX}$ is fixed it produces a first order phase
transition.  The same is true for a fractional threshold:
$r_{sX}(j,k)=0$ for $j<k\alpha$ ; $r_{sX}(k,j)=1$ for $j\geq k\alpha$,
where $\alpha\in (0,1)$. We can construct a continuous approximation for
the fractional threshold $\alpha=1/2$ as
$r_{sX}(j,k)=3(j/k)^2-2(j/k)^3$.  For this $r_{sX}$ and for a single
valued distribution $P_{sX}(k)=\delta_{k\ell}$ the smallest value of
$\ell$ for which there is a first order phase transition is $\ell=8$.

\section*{\it Acknowledgments}

The Boston University work was supported by DTRA Grant HDTRA1-14-1-0017,
by DOE Contract DE-AC07-05Id14517, and by NSF Grants CMMI 1125290, PHY
1505000, and CHE-1213217.  Yeshiva work was also supported by
HDTRA1-14-1-0017. SVB acknowledge the partial support of this
research through the Dr. Bernard W. Gamson Computational
Science Center at Yeshiva College. MAD and LAB wish to thank to UNMdP, FONCyT and CONICET
(Pict 0429/2013, Pict 1407/2014 and PIP 00443/2014) for financial
support. HHAR wish to thanks to FAPEMA (UNIVERSAL 1429/16) for financial
support.

\section*{\it Contributions}

All authors designed the research, analyzed data, discussed results,
and contributed to writing the manuscript. MAD, LDV, SVB and LAB
implemented and performed numerical experiments and simulations. 

\section*{Additional information}

Competing financial interests: The authors declare no competing financial interests.



\end{document}